\documentstyle[11pt]{article}
\def\bd{
\begin{document}} 
\def\ed{\end{document}}
\def\be{\begin{equation}}
\def\ee{\end{equation}}
\def\ba{\begin{array}}
\def\ea{\end{array}}
\def\bea{\begin{eqnarray}}
\def\eea{\end{eqnarray}}
\def\nn{\nonumber}
\let\la=\label
\let\bm=\bibitem 
\def\qq{\quad\quad}

\let\a=\alpha
\let\b=\beta 
\let\c=\gamma
\let\C=\Gamma  
\let\d=\delta
\let\e=\epsilon
\def\eb{{\bar\epsilon}}
\def\tb{{\bar\theta}}
\def\kb{{\bar\kappa}}
\def\etab{{\bar\eta}}
\def\ve{\varepsilon}
\let\z=\zeta 
\let\h=\eta  
\let\k=\kappa
\let\l=\lambda
\let\L=\Lambda
\let\m=\mu
\let\n=\nu 
\let\r=\rho
\def\vt{{\vec\tau}}
\let\s=\sigma
\let\t=\theta
\let\w=\omega 

\def\p{\partial}
\def\pth{(\partial_1+\partial_2+\partial_3)}
\def\ni{\noindent} 
\def\ra{\rightarrow}
\def\lra{\leftrightarrow}

\def\ft#1#2{{\textstyle{{\scriptstyle #1}
\over {\scriptstyle #2}}}}
\def\fft#1#2{{#1 \over #2}}
\def\sst#1{{\scriptscriptstyle #1}}
\newcommand{\eq}[1]{(\ref{#1})}
\def\eqs#1#2{(\ref{#1}-\ref{#2})}
\def\cites#1#2{\cite{#1}-\cite{#2}}
\def\Hat#1{\widehat{#1}}
\def\nosum{({\rm no\ sum\ over}\ i~)}

\def\pl#1#2#3{Phys.~Lett.~{\bf {#1}B} (19{#2}) #3}
\def\np#1#2#3{Nucl.~Phys.~{\bf B{#1}} (19{#2}) #3}
\def\cqg#1#2#3{Class.~and Quant.~Gr.~{\bf {#1}} (19{#2}) #3}
\def\mpl#1#2#3{Mod.~Phys.~Lett.~{\bf A{#1}} (19{#2}) #3}
\def\pr#1#2#3{Phys.~Rev.~{\bf {#1}D} (19{#2}) #3}
\def\prep#1#2#3{Phys.~Rep.~{\bf {#1}C} (19{#2}) #3}

\thispagestyle{empty}

\bd

\begin{titlepage}

\hfill{CTP TAMU-21/97}

\hfill{hep-th/9704057}

\vspace{30pt}

\centerline {\Large\bf Superparticles in $D>11$}

\vspace{30pt}

\centerline{{\large I. Rudychev and E. Sezgin}\footnote{Research
supported in part by NSF Grant PHY-9411543}}

\vspace{15pt}

\centerline{\it Center for Theoretical Physics, Texas A\&M University,}
\centerline{\it College Station, Texas 77843, U.S.A.}

\vspace{50pt}

\centerline{ABSTRACT}

\vspace{15pt}

Actions for two-superparticle system in $(10,2)$ dimensions and
three-superparticle systems in $(11,3)$ dimensions are constructed.
These actions have worldline bosonic and fermionic local symmetries, and
target space global supersymmety generalizing the reparametrization,
$\k$-symmetry and Poincar\'e supersymmetry of the usual superparticle.
With the second particle, or the second and third particles on-shell,
they describe a superparticle propagating in the background of a second
superparticle in $(10,2)$ dimensions, or two other superparticles in
$(11,3)$ dimensions. Symmetries of the action are shown to exist in
presence of super Yang-Mills background as well.

\end{titlepage}


\section{ Introduction }


The possibility of a super $p$-brane in $(10,2)$-dimensions was
conjectured long ago \cite{duff}, in the context of a generalized
brane-scan. More recently, there have been indications for the existence
of a (10,2) dimensional structure in $M$-theory \cite{v,km,b1}.
Motivated by these considerations, super Yang-Mills equations of motion
in (10,2) dimensions were constructed in \cite{ns}. This result has been
recently generalized to describe the equations of motion of supergravity
in (10,2) dimensions \cite{hn}. Previously, possible existence of hidden
symmetries descending from (11,2) dimensions was pointed out \cite{b2}.
Recently \cite{bk2}, it has been suggested that there may be a $(11,3)$
dimensional structure in the master theory, and even the possibility of
a $(12,4)$ dimensional structure has been speculated in \cite{km2}.
These considerations motivated one of the authors to look for an
extension of the work presented in \cite{ns} to higher dimensions, and
it was found that the construction of \cite{ns} generalizes naturally to
$(11,3)$ dimensions. An extension beyond $(11,3)$ dimensions ran into an
obstacle \cite{es1}, which has been removed in \cite{hn2}, where super
Yang- Mills equations have been constructed in $(8+n,n)$ dimensions, for
any $n\ge 1$.

The symmetry algebras realized in the field theoretic models just
mentioned are \cite{b2,ns,es1,b3}: 
\bea
(10,2):\qquad \{ Q_\a, Q_\b\} &=& 
	(\c^{\m\n})_{\a\b}~P_\m~n_\n\ ,\la{an2}\\ 
(11,3):\qquad \{Q_\a, Q_\b\} &=& 
	(\c^{\m\n\r})_{\a\b}~P_\m~n_\n ~m_\r \ , \la{an3}
\eea 
where $n_\m$ and $m_\m$ are mutually orthogonal constant null vectors.
These break the $(10,2)$ or $(11,3)$ dimensional covariance. In order to
maintain this covariance, it is natural to replace the null vectors by
momentum generators \cite{b2} (see also \cite{bk2}), thereby obtaining
\footnote
{
In $(8+n,n)$ dimensions, the full set of generators occuring on the
right hand side of $\{ Q_\a, Q_\b\}$ are $p$-form generators  with
$p=n_0, n_0+4,...,n+4$, where $n_0= n~{\rm mod}~4$. For example, in 
$(17,9)$ dimensions, there are $p$-form generators with $p=1,5,9,13$.
However, actions of the type considered here for $n$-particle systems 
naturally select the $n$\,th rank generator.
}
\bea
(10,2):\qquad \{ Q_\a, Q_\b\} &=& 
	(\c^{\m\n})_{\a\b}~P_{1\m}~P_{2\n}\ ,\la{alg2}\\ 
(11,3):\qquad \{Q_\a, Q_\b\} &=& 
	(\c^{\m\n\r})_{\a\b}~P_{1\m}~P_{2\n}~P_{3\r} \ . \la{alg3}
\eea 
The algebras \eq{an2} and \eq{alg2} first made their appearances in
\cite{b2}, and \eq{an3} and \eq{alg3} in \cite{es1,b3}. In particular,
\eq{alg2} has been put to use in \cite{b1,b2} in the context of higher
dimensional unification of duality symmetries; in \cite{bk1} where four
dimensional bi-local field theoretic realizations are given and
interesting physical consequences such as family unification are
suggested; and in \cite{bk2}, where a two-particle realization in
$(10,2)$ dimensions, in the purely bosonic context, was given. 

The purpose of this paper is to present a supersymmetric extension of
the bosonic two-particle model of \cite{bk2}, and to extend further
these results to (11,3) dimensions, where a three-particle model arises.
We will construct multi-superparticle actions in which the algebras
\eq{alg2} and \eq{alg3} are realized. In doing so, we will find that the
multi-superparticle system has new bosonic local symmetries that generalize
the usual reparametrization \cite{bk2,bk1}, and new fermionic local
symmetries that generalize the usual $\k$-symmetry of the single
superparticle. These symmetries will be shown to exist in
presence of super Yang-Mills background as well. 
 
These results can be viewed as preludes to the constructions of higher
superbranes in (10,2) and (11,3) dimensions. Since the latter should
admit superparticle limits, it is important to develop a better
understanding of the superparticle systems in this context.


\section{Superparticles in (10,2) Dimensions}


\subsection{ A Model with Two Times}


We consider two superparticles which propagate in their respective
superspaces with coordinates $X_i^\m (\tau_1,\tau_2)$ and
$\t_i^\a(\tau_1,\tau_2)$, with $i=1,2$, $\m=0,1,...,11$ and
$\a=1,...,32$. Working in first order formalism, we also introduce the
momentum variables $P^i_\m(\tau_1,\tau_2)$
\footnote{In the earlier version of this paper, we considered a restricted
dependence of variables on proper times (see \eq{rt} and below). We are
grateful to I. Bars for stimulating discussions that led us to consider
more general proper time dependence.}.
The superalgebra \eq{alg2} can be realized in terms of supercharges
\be 
Q_\a=Q_{1\a}+Q_{2\a}\ , 
\ee
with $Q_i(\tau_1,\tau_2)$ defined as 
\footnote{
The spinors are Majorana-Weyl, their indices are chirally
projected, the charge conjugation matrix $C$ is suppressed in
$(\c^{\m_1\cdots \m_p}C)_{\a\b}$, which are symmetric for $p=2,3$ in
$(10,2)$ dimensions, and for $p=3,4$ in $(11,3)$ dimensions. This
symmetry property alternates for $p~{\rm mod}~2$. 
}
\be
Q_{i\a}=\p_{i\a}+\ft14~\c^{\m\n}_{\a\b}~\t_i^\b~P_{1\m}~P_{2\n} \ , 
\quad i=1,2 \ .
\la{q2} 
\ee 
The spinorial derivative is defined as $\p_{i\a}=\p/\p\t_i^\a$, acting
from the right. The transformations generated by the supercharges $Q_\a$
are
\bea
\d_\e X_i^\m = \ft14~\eb\c^{\m\n}(\t_1+\t_2)~\ve_{ij} P_{j\n}\ , 
\qquad  \d_\e \t_i =\e\ , \qquad \d_\e P_i^\m = 0\ , \la{s1} 
\eea
where $\eb\c^{\m\n}\t $ stands for $\e^\a\c_{\a\b}\t^\b$, and $\ve_{ij}$
is the constant Levi-Civita symbol with $\ve_{12}=1$.
Next, it is convenient to define the line element
\be
\Pi_i^\m=(\p_1 +\p_2) X_i^\m -\ft{1}{8} \tb_k \c^{\m\n}(\p_1+\p_2)~
\t_k~\ve_{ij} P_{j\n}\ .  \la{p1}
\ee
While this is not supersymmetric by itself, its product with $P^i_\m$ is
supersymmetric upto a total derivative term, and therefore it is a
convenient building block for an action. The fact that the sum of two
times occur in the line element is a consequence of maintaining
supersymmetry (in the sense just stated) {\it and} the fact that all
field depend on $\tau_1$ and $\tau_2$ (see the end of this section for a
discussion of a restricted time dependence, and its consequences). 

Introducing the symmetric Lagrange multipliers $A_{ij}(\tau_1,\tau_2)$
which are inert under supersymmetry, 
\be
\d_\e A_{ij}=0\ , \la{sa}
\ee
we consider the following action for a two-superparticle system in
$(10,2)$
dimensions
\be
I= \int d\tau_1 d\tau_2 \left(
	P_i^\m \Pi_{i\m} -\ft12 A_{ij} P_i^\m P_{j\m}\right)\ . \la{act1}
\ee  
The action \eq{act1} has a number of interesting symmetries. To begin
with, it is invariant under the target space rigid supersymmetry
transformations \eq{s1}, up to a total derivative term that has been
discarded. Furthermore, it has the local bosonic symmetry 
\be
\d_\L A_{ij} = (\p_1+\p_2)~\L_{ij}\ ,\qquad
\d_\L X_i^\m = \L_{ij} P_j^\m \ , \qquad
\d_\L P_i^\m = 0 \ ,\qquad
\d_\L \t = 0 \ , \la{b1}
\ee
where the transformation parameters have the time dependence
$\L_{ij}(\tau_1,\tau_2)$. Here too, a total derivative term, which has
the form $(\p_1+\p_2) \left(\ft12 \L_{ij} P^i_\m P^{j\m} \right)$, has
been dropped. The diagonal part of these transformations are the usual
reparametrizations, combined with a trivial symmetry of the action
\cite{pkt2}. The off-diagonal part of the symmetry are gauge symmetries
which are the first order form of those which are in the bosonic
two-particle model of \cite{bk2}. Together with the reparametrization
symmetries, they allow us to eliminate the correct amount of degrees of
freedom to yield 8 bosonic physical degrees of freedom for each
particle.

The action \eq{act1} has also local fermionic symmetries which
generalize the usual $\k$-symmetries. Let us denote the $j$th symmetry
of the $i$th particle by $\k_{ij}(\tau_1,\tau_2)$. One finds that the
action \eq{act1} is invariant under the following transformations
\bea
\d_\k \t_i &=& \c^\m \k_{ij} P_{j\m}\ ,\nn\\
\d_\k X_i^\m &=& \ft{1}{4} \left(\tb_k\c^{\m\n}\d_\k \t_k\right)\ve_{ij} 
P_{j\n}\ ,  
\nn\\
\d_\k P_i^\m &=& 0\ ,  
\nn\\
\d_\k A_{ij} &=& \ft12 \kb_{ki}\c^\m (\p_1+\p_2) \t_k~
\ve_{\ell j} P^\ell_\m + (i\lra j)\ .
\la{k1}
\eea
In showing the $\k$-symmetry of the action, it is useful to note the
lemma
\be
P^i_\m (\d_\k \Pi_i^\m) = (\p_1+\p_2) \left( \ft18 \tb_k \c^{\m\n} \d_\k
\t_k \ve_{ij} P^i_\m P^j_\n \right) 
-\ft14 (\d_\k\tb) \c^{\m\n} (\p_1+\p_2) \t_k \ve_{ij} P^i_\m P^j_\n  \ . 
\la{lemma1}
\ee
The diagonal part of the $\k_{ij}$-transformations are the $\k$-symmetry
transformations that resemble the ones for the usual superparticle. The
off-diagonal $\k$-transformations are their generalizations for a
two-superparticle system. Just as the off-diagonal $\L_{ij}$
transformations are needed to obtain 8 bosonic degrees of freedom for
each particle, the off-diagonal $\kappa_{ij}$ symmetries are needed to
obtain 8 fermionic degrees of freedom for each particle. The commutator
of two $\k$-transformations closes on-shell onto the
$\L$-transformations
\be
[\d_{\k_{(1)}},\d_{\k_{(2)}}]= \d_{\L_{(12)}} \ , \la{close}
\ee
where the composite gauge transformation parameter is
\be
\L_{(12){ij}} = \ft12 \kb_{(2)ki}\c^{\m\n} \k_{(1)kj} P_{1\m}P_{2\n}
	 + (i \leftrightarrow j) \ . \la{c1}
\ee
It is clear that the remaining part of the algebra
is $~[\d_\k,\d_\L]=0~$ and $~[\d_{\L_1},\d_{\L_2}]=0~$.

The field equations that follow from the action \eq{act1} take the form 
\bea
&&
P_i^\m P_{j\m}= 0\ ,  \la{e1}\\
&&
\c^{\m\n}  P_{1\m} P_{2\n} (\p_1+\p_2) \t_i= 0\ , \la{e2}\\
&&
(\p_1+\p_2) X_i^\m = \left( A_{ij}\eta^{\m\n} 
+\ft14 \ve_{ij} \tb_k \c^{\m\n} (\p_1+\p_2) \t_k \right) P^j_\n\ ,
\la{e3}\\
&&
(\p_1+\p_2) P_{i\m} = 0\ . \la{e4}
\eea 
While the derivatives occur only in the combination $(\p_1+\p_2)$, the
fields can depend both on $\tau_+$ and $\tau_-$ defined by
$\tau_\pm \equiv \tau_1\pm \tau_2$. It is possible, for example, to restrict
the proper time dependences as follows
\be
X_i^\m (\tau_i)\ , \qq P_i^\m (\tau_i)\ ,\qq \t_i (\tau_i)\ . \la{rt}
\ee
The action still is given by \eq{act1}, with the line element now 
taking the form
\be
\Pi_i^\m=\p_i X_i^\m -\ft{1}{4} \tb_i \c^{\m\n} \p_i~
\t_i~\ve_{ij} P_{j\n}\ .  \la{po}
\ee
It is understood that the free indices of the left hand side are not not
be summed over on the right hand side. The bosonic and fermionic
symmetries discussed earlier retain their forms. In particular, the
global supersymmetry transformations \eq{s1} remain the same, and one
can express the bosonic symmetries \eq{b1} as
\be
\d_\L A_{ij} = \p_i\L_{ij} + \p_j \L_{ji}-\d_{ij}\p_i\L_{ij}\ ,
\qquad
\d_\L X_i^\m = \L_{ij} P_j^\m\ ,\qquad \d_\L P_i^\m =0\ ,
\qquad
\d_\L \t = 0\ , \la{gto}
\ee
where the parameters $\L_{ij}$ depend on $\tau_1$ and $\tau_2$.
Restricting the proper time dependence of the $\k$-symmetry parameter as
$\k_{ij}(\tau_i)$, the fermionic symmetries \eq{k1} can be simplified to
take the form
\bea
\d \t_i &=& \c^\m \k_{ij} P_{j\m}\ ,
\nn\\
\d X_i^\m &=& \ft14 \left(\tb_i\c^{\m\n}\d\t_i\right)\ve_{ij} P_{j\n}\ , 
\nn\\
\d P_i^\m &=& 0\ , 
\nn\\
\d A_{ij} &=& \ft12\left( \kb_{kj}\c^\m \p_k\t_k
+\kb_{ij}\c^\m \p_i\t_i\right)\ve_{ki} P_{k\m} + (i\lra j) \ , \la{kappa1}
\eea
which close on-shell as in \eq{close}, and the composite parameter now
takes the form 
\be
\L_{(12){ij}} = \ft12 \kb_{(2)ii}\c^{\m\n} \k_{(1)ij}P_{1\m}P_{2\n}
	 -(1 \lra 2) \ . \la{close1}
\ee
Similarly, the field equations become
\bea	
&& P_i^\m P_{j\m} = 0 \ , 
\qq 
\p_i P_{i\m} = 0 \ , 
\qq
\c^{\m\n} P_{1\m} P_{2\n} \p_i \t_i= 0 \ , \\
&&
\p_i X_i^\m =\left( A_{ij}\eta^{\m\n} 
+\ft12 \ve_{ij} \tb_i \c^{\m\n} \p_i\t_i \right) P_{j\n} \ .
\eea
Note that the $\k$-symmetry transformation of, say $X_1^\m $, maps a
function of $\tau_1$ to a function of $\tau_1$ {\it and} $\tau_2$. While
this may seem somewhat unusual, it does not present any inconsistency,
and in particular, there is no need to take the momenta to be constants.
The important point to bear in mind is that the symmetry transfomations
close and that they are consistently embedded in a larger set of
transformations that map functions of $(\tau_1,\tau_2)$ to each other. 


\subsection{ Putting the Second Particle On-Shell}


In order to obtain an action for the first particle propagating in the
background of the second particle, we will follow the following
procedure. Recall that $\tau_\pm \equiv \tau_1\pm \tau_2$. Let us also
define $\t_\pm \equiv \t_1\pm \t_2$. We use the $ X_2, A_{22}, \t_- $
equations of motion in the action, thereby putting the second particle
on-shell. However, the remaining fields still have $\tau_-$ dependence.
In analogy with Kaluza-Klein reduction, we then set $\p_-=0$, and
integrating the action over $\tau_-$ to obtain:
\be
I=\int d\tau \left[ P_\m \left(\Pi^\m -A n^\m \right)
-\ft12 e P^\m P_\m \right]\ . \la{action2}
\ee
where the label $1$ has been suppressed throughout, the $\tau_-$
interval is normalized to $1$ and 
\bea
&& 
P_2^\m \equiv n^\m \ , \nn\\
&&
A_{11}\equiv e\ , \quad A_{12}\equiv A \ , \nn\\
&&
\Pi^\m = \p_\tau X^\m-\ft14\tb \c^{\m\n}\p_\tau \t~n_\n\ , \la{np}
\eea
where we have defined $\tau_+\equiv \tau$ and $\t_+ \equiv \t/\sqrt{2}$.
The vector $n^\m$ is constant and null as a consequence of $X_2$ and
$A_{22}$ equation of motion, and the fact that we have set $\p_- =0$. 

The action \eq{action2} is invariant under the local bosonic
transformations
\be
\d e = \p_\tau \xi\ ,\qquad \d A = \p_\tau \L\ , \qquad 
\d X^\m = \xi P^\m +\L n^\m \ ,\qquad \d P^\m=0\ ,
\qquad \d \t =0\ ,
\ee
obtained from \eq{b1} by setting $\p_-=0$ and using the notation $\xi
\equiv \L_{11}$ and $\L \equiv \L_{12}$. The action \eq{action2} is also
invariant under the global supersymmetry transformations \eq{s1}
\be
\d_\e X^\m = \ft14~\eb\c^{\m\n}\t~n_\n \ , 
\qquad  \d_\e\t =\e\ , \qquad \d_\e P^\m = 0\ , \qquad \d_\e e=0\ ,
\qquad \d_\e A=0\ ,
\la{s2} 
\ee
(we have rescaled $\e\ra \e/\sqrt{2}$ for convenience) and invariant
under the local fermionic $\k$ and $\eta$ transformations 
\bea
\d\t &=& \c^\m P_{\m}\kappa + \c^\m n_\m \eta\ ,
\nn\\
\d X^\m &=& \ft14 \tb\c^{\m\n} n_\n \left(\d_\k\t+\d_\eta \t\right)\ ,
\nn\\
\d P^\m &=& 0\ , 
\nn\\
\d e &=& -\kb\c^\m\p_\tau\t~n_\m\ ,
\nn \\
\d A &=& \ft12 \kb \c^\m P_\m \p_\tau \t 
	-\ft12 \etab \c^\m n_\m \p_\tau \t\ , 
\la{k2}
\eea
obtained from \eq{k1} by setting $\p_-=0$, using the field equations of
the second particle, setting $\k_{-i}=0$ , and using the notation
$\k_{+1}\equiv \sqrt{2} \k$ and $\k_{+2}\equiv \sqrt{2}\eta $. The
parameter $\k_{-i}$ has been set equal to zero, as it is associated with
the transformations of $\t_-$ that has been put on-shell, and
which has consequently dropped out in the action.

An alternative way to arrive at the same results is to start from the
restricted model described at the end of Sec. (2.1), again by putting the
second particle on-shell, and this time integrating over $\tau_2$.


\subsection{Introducing Super Yang-Mills Background}


The coupling of Yang-Mills background is best described in the second
order formalism. Elimination of $P^\m$ in \eq{in3} gives  
\be
I_0=  \ft12 \int d\tau~e^{-1} \Pi^\m \left( \Pi_\m -A n_\m\right)\ .
\la{ac1}
\ee      
The bosonic and fermionic symmetries of this action can be read off from
\eq{gt3} and \eq{tr3} by making the substitution $P^\m \rightarrow e^{-1}
(\Pi^\m-A n^\m)$. To couple super Yang-Mills background to this
system, we introduce the fermionic variables $\psi^r,\ r=1,...,32$,
assuming that the gauge group is $SO(32)$. The Yang-Mills coupling can
then be introduced as 
\be
I_1= \int d\tau~\psi^r \p_\tau \psi^s \p_\tau Z^M A_M^{rs} ,\la{ac2}
\ee
where $Z^M$ are the coordinates of the $(10,2|32)$ superspace, and $A_M^{rs}$
is a vector superfield in that superspace. 

The torsion super two-form $T^A=dE^A$ can be read from the superalgebra \eq{alg2}:
\be
T^c = e^\a \wedge e^\b\, (\c^{cd})_{\a\b}\, n_d\ , \quad\quad  T^\a =0\ ,
\la{tc}
\ee
where the basis one-forms defined as $e^A=d Z^M E_M{}^A$ satisfy
\be
d e^c= e^\a \wedge e^\b\ (\c^{cd})_{\a\b}\ n_d\ ,\quad\quad d
e^\a = 0\ , \la{se}
\ee
and $a,b,c,... $ are the (10,2) dimensional tangent space indices.

Using these equations, a fairly standard calculation \cite{w,bst} shows that the
total action $I=I_0+I_1$ is invariant under the fermionic gauge
transformations provided that the Yang-Mills super two-form is given by
\cite{ns}
\be
F=e^\a \wedge e^b \left[\,  n_b \chi_\a - 2(\c_b\l)_\a\, \right]
+ \ft12 e^a \wedge e^b\
F_{b a}\ , \la{fc}
\ee
where we have introduced the chiral spinor superfield $\chi_\a$ and the
anti-chiral spinor superfield $\l$, and that the transformation rules
for $e$ and $A$ pick up the extra contributions \bea
\d_{\rm extra} e &=&  -4e \psi^r\p_\tau \psi^s (\kb \l_{rs})\ , \nn\\
\d_{\rm extra} A &=& \psi^r\p_\tau \psi^s \kb\left( 2\l_{rs}+(\Pi^a-An^a)
	\right)\c_a\chi_{rs}
	+ e \psi^r\p_\tau \psi^s \etab\left( 4\l_{rs}-
	\c_an_a\chi_{rs}\right)\ ,\nn\\
\d_{\rm extra} \psi^r &=& - \d \t^\a~A_\a^{rs} \psi^s\ .
\eea
In addition, the spinor $\l$ must satisfy the condition $ n^a\c_a \l
=0$. One can show that $F$ given in \eq{fc} satisfies the
Bianchi identity $DF=0$, and the constraints on $F$ implied by \eq{fc}
lead to the field equations of the super Yang-Mills system in (10,2)
dimensions \cite{ns}. 


\section{Superparticles in (11,3) Dimensions}


\subsection{A Model with Three Times}


In this section, we consider three superparticles propagating in
$(11,3)$ dimensional spacetime, and take their superspace coordinates to
be $X_i^\m (\vt), \t_i^\a(\vt)$ and momenta $P_i^\m (\vt)\ (i=1,2,3)$,
where $\vt=(\tau_1,\tau_2,\tau_3)$. Following the same reasoning
as in Sec. (2.1), we consider the following action
\be
I= \int d\tau_1 d\tau_2 d\tau_3~\left(
	P_i^\m \Pi_{i\m} -\ft12 A_{ij} P_i^\m P_{j\m}\right)\ , 
\la{act3}
\ee  
where
\be
\Pi_i^\m= \pth X_i^\m -\ft{1}{36} \tb_k \c^{\m\n\r}\pth \t_k~\ve_{ijk} 
P_{j\n}P_{k\r}\ .
\la{pi3}
\ee
The action is invariant under the local bosonic transformations
\be
\d_\L A_{ij} = (\p_1+\p_2+\p_3)~\L_{ij}\ ,\qquad
\d_\L X_i^\m = \L_{ij} P_j^\m \ , \qquad
\d_\L P_i^\m = 0 \ ,\qquad 
\d_\L \t = 0 \ . \la{b3}
\ee
The action is also invariant (modulo discarded total derivative terms)
under the global supersymmetry transformations
\bea
\d_\e X_i^\m = \ft1{12}~\eb\c^{\m\n\r} (\t_1+\t_2+\t_3)~\ve_{ijk}
P_{j\n}P_{k\r}\ , 
\qquad  \d_\e\t_i =\e\ , \qquad \d_\e P_i^\m = 0\ , 
\qquad  \d_\e A_{ij}=0\ ,
\la{susy3} 
\eea
and local fermionic transformations
\bea
\d_\k \t_i &=& \c^\m \k_{ij} P_{j\m}\ ,\nn\\
\d_\k X_i^\m &=& \ft1{12} \left(\tb_k \c^{\m\n\r}\d\t_k \right)\ve_{imn} 
	P^m_\n P^n_\r \ ,\nn\\
\d_\k P_i^\m &=& 0\ , \nn\\
\d_\k A_{ij} &=& \ft16 \kb_{ki}\c^{\m\n} \pth \t_k
	      \ve_{jmn} P^m_\m P^n_\n+ (i\lra j)\ .
\la{k3} 
\eea
The algebra closes on-shell as in \eq{c1}, with the composite gauge
parameter now given by 
\be
\L_{(12){ij}} = \ft13 \kb_{(2)ki}\c^{\m\n\r} \k_{(1)kj}
	        P_{1\m}P_{2\n}P_{3\r} +(i \lra j)\ . 
\la{c2}
\ee
The remaining part of the algebra is $~[\d_\k ,\d_\L]=0~$ and
$~[\d_{\L_1},\d_{\L_2}]=0~$. The equations of motion are similar to
\eq{e1}-\eq{e4}. In fact, all the formulae of this section are very
similar to those for two-superparticles given in Sec. (2.1) and their
$n$ superparticle extension is straightforward.


\subsection{Putting the Second and Third Particles On-Shell} 


To obtain an action which describes the propagation of the first
particle in the background of the other two particles, we follow the
steps described in Sec. (2.2). Let $\tau\equiv \tau_1+\tau_2+\tau_3$,
and $\t\equiv \t_1+\t_2+\t_3$, and denote the orthogonal
combinations by $\tau_\pm$ and $\t_\pm$. Using the equations of motion
for $\t_\pm, X_i, A_{ii}\ (i=2,3)$  and restricting the proper time
dependence of fields by setting $\p_\pm=0$, we obtain the action 
\be
I=\int d\tau \left[ -\ft12 e P^\m P_\m 
+ P_\m \left(\Pi^\m -A n^\m - B m^\m \right)\right]\ , \la{in3}
\ee
where $A\equiv A_{12}$ and $B\equiv A_{13}$ and
\be
\Pi^\m = \p_\tau X^\m-\ft16\tb \c^{\m\n\r}\p_\tau \t~n_\n m_\r\ .
\la{pin}
\ee
The vectors $n^\m$ and $m_\m$ are mutually orthogonal constant null
vectors, as a consequence of $A_{ii}$ and $X_i^\m$ equations of motion
for $i=2,3$ resulting from the original action \eq{act3}, and having set
$\p_\pm =0$. The action \eq{in3} is invariant under the bosonic
transformations
\be
\d e = \p_\tau \xi\ ,\qquad \d A = \p_\tau \L\ , \qquad 
	\d B = \p_\tau \Sigma\ , \qquad
\d X^\m = \xi P^\m +\L n^\m + \Sigma m^\m \ , \la{gt3}
\ee
where $\xi\equiv \L_{11}$, $\L \equiv \L_{12}$ and $\Sigma\equiv \L_{13}$.
The action is also invariant under the global supersymmetry
transformations
\be
\d_\e X^\m = \ft1{12}~\eb\c^{\m\n\r} \t~n_\n m_\r \ , 
\qquad  \d_\e \t =\e\ , \qquad \d_\e P^\m = 0\ , 
\qquad \d_e A=0\ , \qquad \d_e B=0\ , \la{isusy3} 
\ee
and local fermionic $\k$, $\eta$ and $\omega$ transformations 
\bea
\d\t &=& \c^\m P_{\m}\kappa + \c^\m n_\m \eta
	+ \c^\m m_\m \omega\ , \nn\\
\d X^\m &=& \ft16 \tb\c^{\m\n\r} n_\n m_\r \left(\d_\kappa \t+\d_\eta\t 
	+\d_\omega \t\right)\ , \nn\\
\d P^\m &=& 0\ , \nn\\
\d e &=& -\ft23 \kb \c^{\m\n} \p_\tau \t n_\m m_\n\ , \nn \\
\d A &=&  -\ft13 \kb \c^{\m\n} \p_\tau \t m_\m P_\n 
	-\ft13 \etab \c^{\m\n} \p_\tau \t n_\m m_\n\ , \nn \\
\d B &=&  -\ft13 \kb \c^{\m\n} \p_\tau \t P_\m n_\n 
	-\ft13 \kb \c^{\m\n} \p_\tau \t n_\m m_\n
	-\ft13 {\bar \omega}\c^{\m\n} \p_\tau \t n_\m m_\n \ , \la{tr3} 
\eea
where $\k, \eta,\w$ are equivalent to $\k_{+i} (i=1,2,3)$ upto a
constant rescaling, and the irrelevant parameters $\k_{\pm i}$ have been
set equal to zero.


\section {Conclusions}


We have presented simple action formulae for two- and
three-superparticle systems in $(10,2)$ and $(11,3)$ dimensions,
respectively. The symmetries of the action exhibit interesting
generalizations of reparametrization and $\k$-symmetries. An action
similar to \eq{ac1}-\eq{ac2} can easily be constructed for the $(11,3)$
dimensional superparticle and it implies the $(11,3)$ dimensional super
Yang-Mills equations \cite{rss}. We also expect that the action
\eq{ac1}-\eq{ac2} can be generalized to obtain a heterotic string action
and possibly other superbrane actions. 

The $n$ particle models constructed here ($n=2,3)$ make use of $n$
fermionic coordinates $\t_i\ (i=1,...,n)$. The fact that they all
transform by the same constant parameter $\ve$ suggests that we can
identify them: $\t_i=\t$. It is also natural from the group theoretical
point of view to associate the coordinates $X_i^\m, \t_\a$ with the
generators $P_i^\m, Q_\a$. However, it is not necessary to do so, since
there are sufficiently many local fermionic symmetries to give $8$
physical fermionic degrees of freedom for each $\t_i$ (see \eq{k1}).
Thus, it does not seem to be crucial to have one or many fermionic
variables. 

Another, and possibly more significant, feature of the models
constructed here is that they involve multi-times, in the sense that
fields depend on $\tau_i\ (i=1,...,n)$ over which the action is
integrated over. The derivatives occuring in the action come out to be
the sum of these times, which indicates that any (pseudo) rotational
symmetry among them is lost. One might therefore be tempted to declare
the fields to depend on the single time $\tau=\tau_1+\cdots \tau_n$.
Perhaps this is the sensible thing to do, however, we have kept the
multi-time dependence here, partically motivated by the fact that our
results may give a clue for the construction of an action that involves
an $(n,n)$ dimensional worldvolume. If an action can indeed be
constructed for $(n,n)$ brane, i.e. brane with an $(n,n)$ dimensional
worldvolume, one may envisage a `particle limit' in which the spatial
dependence is set equal to zero, yielding an $(n,0)$ dimensional
worldvolume. If the worldvolume diffeomorphisms are to be maintained,
then we may need to consider a term of the form $P^{ia}_\m \p_a X_i^\m$
in the action, where $a=1,...,n$ labels the $(n,0)$ worldvolume
coordinates. It is not clear, however, how supersymmetry and
$\k$-symmetry can be achieved in this setting.

Notwithstanding these open problems, one may proceed to view the
coordinates $(\tau_1,\sigma_1)$ and $(\tau_2,\sigma_2)$ as forming a
$(2,2)$ dimensional worldvolume embedded in $(10,2)$ dimensions, in the
case of a two-superstring system. Similarly, a three-superstring system
would form a $(3,3)$ dimensional worldvolume embedded in $(11,3)$
dimensions. 

It is of considerable interest to construct a string theory in $(3,3)$
dimensions which would provide a worldvolume for a $(3,3)$ superbrane
propagating in $(11,3)$ dimensions \cite{es1}, thereby extending the
construction of \cite{v,km} a step further \cite{rss}. Indeed, a string
theory in $(n,n)$ (target space) dimensions has been recently constructed 
\cite{rss}. This theory is based on an $N=2$ superconformal algebra for the
right-movers in $(n,n)$ dimensions, and an $N=1$ superconformal algebra
for the left-movers in (8+n,n) dimensions. The realizations of these
algebras contain suitable number of constant null vectors, which arise
in the expected manner in the algebra of supercharge vertex operators
\cite{km2}. Furthermore, the massless states are expected to assemble
themselves into super Yang-Mills multiplet in $(8+n,n)$ dimensions. Much
remains to be done towards a better understanding of this theory, and
its implications for the target space field equations that may exhibit
interesting geometrical structures that generalize the self-dual
Yang-Mills and gravity equations in $(2,2)$ dimensions \cite{hull3}. 

Further studies of supersymmetry in $D>11$ may also be motivated by the
fact that they contain both the type IIA and IIB supersymmetries of ten
dimensional strings \cite{b3,rss}. Therefore, it would be interesting to
find a brane-theoretic realization of the $D>11$ symmetries that would
provide a unified framework for the description of all superstrings in
$(9,1)$ dimensions. 

\bigskip


\noindent{\bf Acknowledments}


\bigskip

We dedicate this article to the memory of D.V. Volkov who is one of the
fathers of supersymmetry and whose original work on the theory of
superparticles and superbranes has been very inspirational to us. We are
grateful to I. Bars, C.S. Chu and R. Percacci for many helpful
discussions. We also thank Z. Khviengia for a useful discussion
concerning the structure of the three-particle action and V. Dobrev for
pointing out reference \cite{db} to us. This work was suppoted in part
by the National Science Foundation, under grant PHY-9722090. 

\bigskip

The following Note Added expands the Note Added that appeared in the
previous version of this paper, in order to explain in more detail the
revisions that were made in the first version of the paper and its
relation with the papers of Bars and Deliduman \cite{deli1,deli2}. This
version of the Note Added is to appear as an Addendum in Phys. Lett. B.

\bigskip


\noindent{\bf  Note Added } 


\bigskip

In the original version of this paper, we proposed the action (10), with
the line element as given in (21), and time dependences of the variables
as in (20). We gave the $\kappa$-symmetry transformations (23), the
supersymmetry transformations (7), the bosonic symmetry transformations
(22) (with some minor errors corrected here) and the symmetry algebra
(14), again with the restricted time dependences (21) understood. The
original version also contained the correct results for the description
of one superparticle in the background of one or two other
superparticles with constant momenta, as well as a super Yang-Mills
background.

Subsequently, Ref. [21] appeared in which an action for a superparticle
in the backcground of a second superparticle with constant momentum was
also constructed. The single time formalism was introduced in [21].
Allowing gauge transformations that depend on the equations of motion,
the results of [21] agree with ours. 

Eventually, in Ref. [22], these models were improved considerably and an
action was proposed for multi-superparticles in which the variables
depend on single time $\tau$, and the momenta are not taken to be
constants.

After the appearance of [22], and motivated by discussions with I. Bars
about the property of the symmetry transformations of our original
model, in which a function of one time is mapped into a function of
another time, we revisited the original version of our paper and
considered a multi-time dependence for all the variables, to see if we
could relax this property. Performing the Noether procedure, we found
that the line elements had to be defined as in (8), and the symmetries
as in (11) and (12). However, observing that: a) only the derivative
with respect to the total time occurs in the action, and b) the index on
the fermionic coordinates, and the first index on the $\kappa$-symmetry
parameter can be viewed as an extended supersymmetry index (denoted by A
in [22]), it follows that the model thus obtained is in effect that of
Bars and Deliduman [22]. 

It should be noted that the multi-time approach: a) enables one to see
the connection between the the single time model of Bars and Deliduman
[22] and our original model with the restricted time dependence of
variables and b) shows that the symmetry transformations that map a
function of one time into a function of another time can consistently be
embedded into a larger set of transformations that map functions of both
times into each other (see below (26) for further comments about this,
and the third paragraph in the conclusions for a further discussion of
the multi-time approach). 

We also note that, focusing on the case of simple supersymmetry, while
the $\kappa$-transformation rules of Ref. [22] contained one
$\kappa$-symmetry parameter (though the existence of $n$ fermionic first
class constraints was observed), we found the explicit form of the
$\kappa$-symmetry transformations that involve $n$ such parameters. A
detailed account of the relation between the parameters, and the gauge
transformations proportional to the field equations that explain the
different forms of the transformation rules is given in [23]. 


\baselineskip=14pt

\bigskip\bigskip


\end{document}